
\magnification=\magstep1
\vsize= 8 true in
\hsize= 6.5 true in
\centerline{\bf A NEW FORMULA FOR THE MASS OF A STATIONARY}

\centerline {\bf  AXISYMMETRIC CONFIGURATION}
\bigskip
\bigskip
\centerline{R.M. Avakian and G. Oganessyan$^*$}
\bigskip
\centerline{Dept. of Theoretical Physics,}
\centerline{Yerevan State University,}
\centerline{375049 Yerevan, Armenia.}
\vskip 5 true in
$^*$ Present address: Theoretical Astrophysics Group, Tata
Institute of Fundamental Research, 400 005 Bombay, India.
E-mail : gurgen@tifrvax.tifr.res.in
\vfill
\eject

\centerline{\bf ABSTRACT}
\bigskip
\bigskip
  It has been shown by one of the authors$^1$ that in isotropic spherical
  coordinates there is a relation between the mass of a static spherical
  gravitating body and the pressure distribution inside it. In
this paper the result  is generalized for the case of stationary
axisymmetric configurations.

\vfill
\eject

{\bf1}. It has been shown that in the case of a static spherically symmetric
distribution of matter there exists the following relation$^1$

$$  M^2 = {{32} \pi\over {k}} \int_{0}^{r_s}P(r)r^3 e^{{\nu +3\lambda}\over
2} dr , \eqno (1)$$

     where $M$ is the mass of the spherical configuration, $P(r)$   is the
pressure distribution inside, $k$ is the gravitational constant
and $\nu$ and $\lambda$ are the metric functions, with
$r_s$ being the boundary of the configuration, given by $P(r_s)=0$.
 It should be mentioned that $r$ is the isotropic radial
	coordinate, i.e. the metric is written in the form

$$ ds^2 = e^{\nu(r)} c^2 dt^2 - e^{\lambda(r)}[ dr^2 + r^2 ( d\theta^2
+ \sin^2\theta d\phi^2)] . \eqno (2) $$

Taking the Newtonian limit of (1) one gets
$$  M^2 = {32 \pi\over{k}}\int_{0}^{r_s}P(r)r^3  dr . \eqno (3)$$
The same relation can be easily obtained in the framework of Newtonian
theory$^1$.

{\bf2}. Now we shall try to generalize the formula (1) for the case of
 axisymmetric gravitational fields, which can be produced either by
 a stationarily rotating axisymmetric configuration or by a motionless
 body with a similar distribution of matter. In the latter case
one should assume the presence of internal stresses in the matter.

Since (1) can be derived only in isotropic coordinates, the generalization
is possible in the coordinates which in the spherically
symmetric limit, for instance when the angular
velocity $\Omega$ becomes zero, go over into the isotropic form.
 It is easy to see that the metric will meet this requirement if written
in the form

$$ ds^2 = ( e^\nu - \omega^2r^2\sin^2\theta e^\mu)c^2 dt^2 - e^\lambda
(dr^2 + r^2  d\theta^2 ) - r^2\sin^2\theta e^\mu d\phi^2 -
2\omega r^2\sin^2\theta e^\mu c d\phi dt . \eqno(4)$$

where $\nu,\mu,\lambda$ and $\omega$  are functions
of $r$ ,$\theta$ and $\Omega$. When
$\Omega = 0$ the distribution is spherical,
 $\nu,\lambda$ and $\mu$ depend only on $r$ and as a
consequence of the spherical symmetry $e^{\lambda}$
and $e^{\mu}$  equate
to each other, so as to make the angular term proportional to
$(d\theta^2 + \sin^2\theta d\phi^2)$.

  It is known$^2$ that if the components of the metric tensor do
not depend on $x^0 = c t$  the component
 $R_0^0$ of the Ricci tensor can be written as
	  $$R_0^0 = {1 \over \sqrt{-g} } {\partial{} \over \partial{x^\alpha}}
	   (\sqrt{-g}\, g^{0 i} \Gamma_{0i}^\alpha)  ,\eqno(5)$$

where $ g= det\, g_{i k} = - r^4 \sin^2\theta e^{\nu + 2\lambda +\mu},
 \,i= 0,1,2,3,\, \alpha =1,2,3$.
 In the axially symmetric case the components of the metric
tensor are independent also of $x^3 =\phi$ , and
the component $R_3^3$  can be written in the same form:

	  $$R_3^3 = {1\over\sqrt{-g} } {\partial{}\over\partial {x^\alpha}}
	   (\sqrt{-g}\,g^{3 i} \Gamma_{3i}^\alpha) . \eqno(6)$$
	    Now we take into account the Einstein equations and consider the
 combination $R_0^0 + R_3^3$
$${1\over\sqrt{-g}} {\partial{}\over\partial{x^\alpha}} [\sqrt{-g}\,(g^{0i}
\Gamma_{0i}^\alpha + g^{3i}\Gamma_{3i}^\alpha ) ] = -{{8\pi k}\over
 c^4} (T_1^1 + T_2^2 ) . \eqno(7)$$

where $T_1^1$ and $T_2^2$  are components of the energy-momentum
tensor which in the case of perfect fluid is given by

$$T_i^k = (P + \rho c^2 ) u_i u^k - P \delta_i^k . \eqno(8)$$

 Calculating $\Gamma_{0i}^\alpha$ and
$\Gamma_{3i}^\alpha$ and feeding them into (7), one gets

$${{\sin\theta}\over r} {\partial{} \over\partial {r}}\,[\,r^3{\partial{} \over
\partial {r}} (e^{{\nu + \mu}\over2}) + {1\over \sin\theta} {\partial
\over \partial \theta}
[\sin^2\theta {\partial \over \partial{\theta}} (e^{{\nu
+\mu}\over 2} )\,] =
 -{{8\pi k}\over {c^4}} ( T_1^1 + T_2^2 ) \sqrt{-g} . \eqno(9)$$

Multiplying both sides by $r \sin\theta$
 and integrating over the whole 3-dimensional space we obtain

          $$\eqalignno{&\int_{0}^{\infty} \int_{0}^{\pi} \int_{0}^{2\pi}
		   \sin^2\theta
           {\partial{} \over \partial{r}} [ r^3{\partial{}\over \partial {r}}
           (e^{{\nu + \mu}\over 2 } )]\, dr d\theta d\phi + \int_{0}^{\infty}
           \int_{0}^{\pi} \int_{0}^{2\pi} {\partial{} \over \partial{\theta}}
           [r\sin^2\theta {\partial{} \over \partial{\theta}} (e^{{\nu +
\mu}\over{2} } )]
           \, dr d\theta d\phi \cr
           &\qquad = - {{8\pi k}\over c^4} \int \int_{V} \int ( T_1^1 + T_2^2 )
           r \sin\theta \sqrt{-g} dr d\theta d\phi .&(10)\cr}$$
The integration in the RHS  of (10) is over the volume V of the body,
 since $T_i^k = 0$ outside the matter.

One can easily take the second integral in the LHS of (10) with respect
to $\theta$   and see that it is zero. Taking
the first integral  with respect to $r$ one gets
$$\eqalignno{&\int_{0}^{2\pi} \int_{0}^{\pi} \sin^2\theta\, d\theta d\phi \,r^3
{\partial{}\over \partial{r}} (e^{{\nu + \mu}\over{2} })
{\Big\vert}_{0}^{\infty}= 2\pi \int_{0}^{\pi} \sin^\theta d\theta
[r^3 {\partial{}\over \partial {r}} (e^{{\nu + \mu }\over{2} }) ]_
{r\rightarrow\infty} \cr
 &\qquad= - {{8\pi k}\over c^4}\int\int_{V}\int (T_1^1 + T_2^2)
r\sin^2\theta dr d\theta d\phi .&(11)\cr}$$

It is notable that the expansion of $e^{\nu + \mu}\over2$  in terms
of  $1\over r$  should start from the $1\over r^2$ term, otherwise the
 integral
will diverge.

In order to calculate the integral in (11) one has to know the expansions
of $e^\nu$ and $e^\mu$ up to  $1\over r^2$
 order for the metric (4) in the external domein. One can start from
 the known expansions in the harmonic coordinates$^3$

 $$\eqalignno{&ds^2 = (1 - {{r_g}\over {R}} + {{r_g}^2\over {2R^2}} ) c^2 dt^2
 - (1 + {r_g\over{R}} + {{r_g}^2\over{2R^2}} ) dR^2 - R^2 (1 +
{r_g\over{R}} + {{r_g}\over{4R^2}} ) (d\theta^2
+ \sin^2\theta d\phi^2 ) \cr
 &\qquad+ 2 {{2kJ}\over{c^2 R}} d\phi dt,&(12)\cr}$$
where $J$ is the angular momentum, $r_g$
the gravitational radius and $M$ the mass of the body.
 One can see that in this approximation the metric coefficients
do not depend on the angular coordinates. Thus, the transition from
(12) to the form (4) can be made by a scale transformation

$$R = r ( 1 + {C\over{r}} + {D\over {r^2}} ) ,\eqno(13)$$

where C and D are unknown constants. Inserting (13) into (12) and
demanding that (12) go over into the form (4), we get $C = 0$, $D
={ {r_g}^2\over 8}$.

 Now we can easily find the metric coefficients written
in $1\over r^2$ approximation in the "isotropic" coordinates

$$e^\nu = 1 -{{ r_g}\over{r}} + {{r_g}^2\over 2r^2} , $$
$$e^\lambda = 1 +{{r_g}\over{r}} + {{3{ r_g}^2}\over8r^2} ,$$
$$e^{{\nu + \mu}\over2} = 1 - {{r_g}^2\over 16r^2 } .\eqno(14)$$

Inserting (14) into (11) and integrating with respect to
 we obtain the formula we have been after:

$$M^2 = - {64\over{k}} \int_{0}^{\pi}\int_{0}^{{r_s}(\theta}
(T_1^1 + T_2^2) \,r^3 \sin^2\theta e^{{\nu + 2\lambda + \mu}\over2}
 dr d\theta , \eqno(15)$$

where ${r_s}(\theta)$ is the boundary of the configuration,
 $P[{r_s}(\theta)] = 0 .$

In the static case, when $P$, $\nu$ and $\lambda = \mu$
 do not depend on the angular coordinate, an elementary
integration with respect to $\theta$ immediately leads to (1).

\vskip 1 cm
{\bf Aknowledgments.}

We would like to thank the participants of the theoretical
seminar at the Yerevan State University and the members of the Theoretical
Astrophysics Group, Tata Institute of Fundamental Research, for useful
discussions.
\vskip 1 cm
{\bf References}

1. Avakian R.M.,{\it Astrofizika},{\bf33},429,1990.

2. L.D. Landau and E.M. Lifshitz,{\it Classical Theory of Fields} (Pergamon,
 New York ,1975).

3. G.S.Saakyan,{\it Space-Time and Gravitation} (Yerevan State
University,1982, in Russian).

\end